\documentclass{article}

\usepackage{arxiv}

\usepackage[utf8]{inputenc} 
\usepackage[T1]{fontenc}    
\usepackage{url}            
\usepackage{booktabs}       
\usepackage{amsfonts}       
\usepackage{microtype}      
\usepackage{graphicx}
\usepackage[nohyperlinks, printonlyused, nolist]{acronym}
\graphicspath{ {./images/} }
\usepackage[%
    style=ieee,%
    maxnames=4,%
    minnames=2,%
    backend=biber,%
	sorting=none,%
	natbib=true,%
	defernumbers=true,%
]{biblatex}%
\DefineBibliographyStrings{english}
   {andothers = \mkbibemph{et al\adddot}}
\bibliography{%
references.bib
}
\def\BibTeX{{\rm B\kern-.05em{\sc i\kern-.025em b}\kern-.08em
    T\kern-.1667em\lower.7ex\hbox{E}\kern-.125emX}}

    \begin{acronym}

    \acro{iot}[IoT]{Internet of Things}
    \acro{mno}[MNO]{Mobile Network Operators}
    \acro{urllc}[uRLLC]{Ultra Reliable and Low Latency Communications}
    \acro{cpps}[CPPS]{cyber-physical production systems}
    \acro{qos}[QoS]{Quality of Service}
    \acro{nm}[NetEm]{network emulation}
\end{acronym}

\title{6G Underlayer Network Concepts for Ultra Reliable and Low Latency Communication in Manufacturing}

\author{
 Daniel Lindenschmitt \\
  Institute for Wireless Communication and Navigation\\
  RPTU Kaiserslautern-Landau\\
  D-67663 Kaiserslautern \\
  \texttt{daniel.lindenschmitt@rptu.de} \\
   \And
 Jan Mertes \\
  Institute for Manufacturing Technology and Production Systems\\
  RPTU Kaiserslautern-Landau\\
  D-67663 Kaiserslautern \\
  \texttt{jan.mertes@rptu.de} \\
  \And
 Christian Schellenberger \\
  Institute for Wireless Communication and Navigation\\
  RPTU Kaiserslautern-Landau\\
  D-67663 Kaiserslautern \\
  \texttt{christian.schellenberger@rptu.de} \\
\AND
 Marius Schmitz \\
  Institute for Manufacturing Technology and Production Systems\\
  RPTU Kaiserslautern-Landau\\
  D-67663 Kaiserslautern \\
  \texttt{marius.schmitz@rptu.de} \\
    \And
 Bin Han \\
  Institute for Wireless Communication and Navigation\\
  RPTU Kaiserslautern-Landau\\
  D-67663 Kaiserslautern \\
  \texttt{bin.han@rptu.de} \\
 \And
 Jan C. Aurich \\
  Institute for Manufacturing Technology and Production Systems\\
  RPTU Kaiserslautern-Landau\\
  D-67663 Kaiserslautern \\
  \texttt{fkb@mv.rptu.de} \\
  \And
 Hans D. Schotten \\
  Institute for Wireless Communication and Navigation\\
  RPTU Kaiserslautern-Landau\\
  D-67663 Kaiserslautern \\
  \texttt{schotten@rptu.de} \\
}

\begin{document}
\maketitle
\begin{abstract}
Underlayer networks in the context of 6G for manufacturing are crucial. They address the evolving needs of highly interconnected and autonomous systems in  industry. The digitalization of manufacturing processes, driven by the Internet of Things and increased data availability, enables more efficient and demand-driven production. However, wireless connectivity, which offers flexibility and easy integration of components, comes with challenges such as signal interference or high latency. A new management system is needed to coordinate and route traffic of multiple networks in a specific coverage area. This paper proposes underlayer networks designed for manufacturing, providing low latency, reliability, and security. These networks enable wireless connectivity and integration of wireless technologies into the manufacturing environment, enhancing flexibility and efficiency. The paper also discusses network slicing, spectrum sharing, and the limitations of current wireless networks in manufacturing. It introduces a network concept for underlayer networks and evaluates its application in closed-loop communication for machine tools. The study concludes with future research prospects in this area.
\end{abstract}

\keywords{Underlayer networks, 6G, manufacturing, network slicing, spectrum sharing, uRLLC, closed-loop communication, network management}

\section{Introduction}
Underlayer networks play a crucial role in the context of 6G for manufacturing as they address the evolving needs of highly interconnected and autonomous systems in the industry. The advancing digitalization of manufacturing processes has opened up new possibilities and flexible concepts, driven by the \ac{iot} and the availability of a greater volume of data in shorter time-frames. This data influx enables more efficient and demand-driven production, with the concept of batch size 1 emerging as a key driver for maximum flexibility and adaptability in manufacturing.

For a successful transition from traditional production facilities to data-driven and autonomous \ac{cpps}, the interconnection of components is essential. Historically, wired solutions were preferred for this purpose. However, the shift towards wireless connectivity of all manufacturing process components has become a critical aspect of the conversion process. Wireless data transmission between different parts of a production facility allow a more flexible system design and easier integration of new components, enabling manufacturers to rapidly adapt to changing market demands and production requirements.

While wireless connectivity brings numerous benefits, there are also related challenges that need to be considered. Compared to wired transmission, wireless transmission is generally less robust and can be subject to interference, signal attenuation, and environmental factors. Additionally, wireless transmission often incurs higher latencies, which can be a limitation for real-time and safety-critical applications that require near-instantaneous response time.

The implementation of the 5G standard has introduced a multitude of innovative features to radio networks, including the emergence of private networks. This progress has not only expanded the reach of 5G technology to conventional \ac{mno} and telecommunication companies but has also made it accessible to any interested businesses. In accordance with the standard, national regulatory authorities are able to define a specific frequency range, which is exclusively reserved and allocated for private networks, strictly prohibiting its nationwide usage by telecommunication companies. Governments and regulatory authorities have the flexibility to govern private networks in a unique manner.

This is where underlayer networks for 6G in manufacturing come into play. These networks are designed to address the specific requirements of the industry, including ultra-low latency, high bandwidth, reliability and security. By providing a robust and efficient communication infrastructure, underlayer networks enable wireless connectivity between different components of a manufacturing process while ensuring that safety-critical applications can be supported. This allows for the seamless integration of wireless technologies into the manufacturing environment, enhancing flexibility, adaptability and efficiency. By bringing the idea of private networks and the concept of underlayer networks together, a new management system is needed which is able to coordinate and route the traffic of multiple networks in a specific coverage area. 

In Section \ref{relwork} we sum up current research topics in the area of network slicing and spectrum sharing for 5G systems and a following 6G standard. Further we specify, why a new underlayer network concept is not possible with current technologies. Additionally we illuminate the gap of current 5G systems with respect to robust and low-latency communication from the point of view of a \ac{cpps}. Afterwards in Section \ref{sec3} we introduce a network concept for manufacturing which is based on underlayer networks and their efficient management. We implement this concept for closed-loop communication for machine tolls in Section \ref{sec4} and evaluate new purposes with respect to \ac{urllc}, followed by Section \ref{concl} with a conclusion and outlook on future work in this area.

\section{Related Work}
\label{relwork}

\subsection{Network Slicing \& Spectrum Sharing}
Network slicing, a key technology in 5G networks, has gained significant attention due to its ability to create virtual networks with customized characteristics to meet specific application requirements. 

In \cite{Slicing1} the authors provide an overview of network slicing in 5G networks, specifically highlighting its application in underlayer networks and discuss the challenges, benefits and potential use cases. The focus of \cite{Slicing2} is on network slicing in underlayer networks for 5G. An analysis of the architecture and implementation challenges of network slicing is presented. Furthermore, a framework for efficient and scalable network slicing in this context is proposed. As a comprehensive survey, the authors of \cite{Slicing3} focus on the enabling technologies for network slicing in 5G networks, with a particular emphasis on their application in underlayer networks. They cover various aspects, including network slicing architectures, resource management, security and service orchestration. In \cite{Slicing4} the authors investigate resource allocation techniques for network slicing in 5G underlayer \ac{iot} networks. They address the challenges of efficient resource allocation, propose algorithms and approaches to optimize resource utilization and enhance the performance of network slicing in this context.

While network slicing is trying to fulfill all demands of an \ac{mno} in the virtual configuration domain on the same hardware resources, spectrum sharing is another possible way of shaping private mobile networks to the desired functionalities on the physical connection by using different hardware resources. Appropriate mechanisms in the spectrum sharing domain can increase spectral efficiency and thus increase the overall capacity of a wireless network. Additionally, each owner of a part of the spectrum in the area of private networks is able to configure the underlayer networks with respect to the demands.

In \cite{Sharing1} the authors discuss the significance of spectrum efficiency in 5G networks and highlight the use of advanced spectrum sharing techniques to improve it. The survey focuses on cognitive radio, device-to-device communication, in-band full-duplex communication, non-orthogonal multiple access and Long Term Evolution on unlicensed spectrum, providing an overview of their principles and research methodologies. The challenges of deploying these techniques in the context of evolving 5G networks are addressed, along with the integration of multiple spectrum sharing techniques and potential challenges. Regulatory aspects of spectrum sharing are also an important topic in this area. The authors of \cite{Sharing3} and \cite{Sharing2} highlight the increasing need for practical solutions to effectively share spectrum bands in next-generation wireless networks. They review various spectrum sharing methods and categorize them based on their operational frequency regime (licensed or unlicensed bands). They also explore potential implementation scenarios and necessary amendments for legacy cellular networks. The paper also discusses the applications of artificial intelligence and machine learning techniques in facilitating spectrum sharing and identifies open research challenges for future investigations.

Spectrum sharing in an upcoming 6G standard can be a promising method to optimize spectrum utilization and might be able to establish an efficient way of implementing underlayer networks. Developing effective management models and considering regulatory aspects are crucial for the success of this technology in a future 6G communication system. In Section \ref{sec3} an underlayer network concept is introduced, which is establishing robust and low-latency communication via a network management system.

\subsection{Communication Technologies in Manufacturing}
\label{sec23}
\ac{cpps} are characterized by high flexibility, scalability and reconfigurability. Therefore, a high degree of interconnected entities are needed to enable decentralized and distributed computing units \cite{mono16}.

Depending on the hierarchical level of the information exchange from enterprise to field device level as defined in ISO IEC 62264-1:2013 \cite{ISO.2013}, different requirements regarding the communication technology exist. Enterprise-level data transmission requires many connected devices as well as high data rate and data integrity. This is usually not subject to strict limitations in terms of latency and reliability. However, for low-level field device communication and real-time closed-loop process control, a reliable and low-latency communication is required \cite{acet19}. Especially for factory automation on machine tool level, low latencies between 0.5 - 10 ms, high reliability with a packet loss rate of $10^{-9}$ \cite{schu17} and time determinism \cite{woll17} is needed. 

Due to that, currently mainly wired technologies (e.g. fieldbus and Ethernet-based systems) are deployed for use cases with real-time requirements \cite{brah15}. However, to meet the required scalability and flexibility for \ac{cpps}, wireless communication networks are needed \cite{gung09}.  

Next to the combination of different, heterogeneous solutions for industrial wireless networks to support different industrial use cases \cite{Li.2017}, mobile communication networks are especially important for automation in manufacturing due to the easy deployment, scalability. Moreover, unlike other wireless radio solutions, cellular networks can meet different communication requirements for different use cases simultaneously \cite{nava20}.

In particular, the development of the 5G mobile communications standard with focus on industrial applications is intended to provide the wireless infrastructure for \ac{cpps}. 

However, currently no industrial wireless network - including currently available private 5G networks with Release 15 \cite{9861872} - can meet the requirements for closed-loop control of machine tools. Due to that, dedicated and often proprietary wireless networks have to be used to enable a range of use cases in manufacturing. In order to maintain the advantage of centralized, wireless administration, the network should be implemented as a underlayer network, which can be administered from a higher-level network and enables seamless data transmission across different networks. An architecture for underlayer networks that meets the requirements for \ac{urllc} and enables wireless closed-loop control of a machine tool is presented in the following section.  

\section{6G underlayer network concepts for manufacturing}
\label{sec3}
In order to increase productivity and reduce costs, it will be necessary to establish an efficient way of data communication in manufacturing systems of the future, which is tailored to the needs of the respective application and still ensures connectivity between all components. By establishing the concept of 6G underlayer networks, we provide a solution for interconnection and coordination of various devices and systems within a factory. They serve as the backbone that enables real-time data exchange, synchronization and collaboration between industrial robots, machine tools, sensors, control systems and other smart machines. 

\subsection{General approach}
Since the introduction of the 5G mobile communications standard and the associated possibility of being able to license part of the frequency spectrum, e.g. in Germany in the range from 3.7 GHz to 3.8 GHz, as private spectrum independently of the \ac{mno}s, a large number of new possibilities and applications have emerged. Their requirements differ from public cellular networks in regards to e.g. network control or the integration of additional functionalities. Network slicing has created the first opportunities in 5G to adapt mobile networks more easily to different requirements and spectrum sharing is already being used to divide public and private networks.

In a future 6G standard, the requirements for flexibility and simpler operation will increase further, especially with regard to private networks, which is why existing technologies will have to be adapted. By establishing so-called underlayer networks, it will be possible to operate a very small-granular network structure that is also capable of ensuring communication with an overlaying network. Underlayer networks are independent cellular communication networks that can be adapted to the respective requirements of the application and, compared to the use of non-cellular standards such as WiFi, offer significant advantages in the area of robust and low-latency communication due to a significantly higher degree of determinism in the network. Independently, other communication standards can be used if they meet the specifications of the application in terms of \ac{qos} parameters as underlayer networks. This ensures parallel operation of cellular and non-cellular networks in different frequency ranges, which can exchange data via a corresponding gateway if required.
Due to the need for networks with different configurations and \ac{qos} requirements or different private mobile networks by more than one operators in the identical area, it is necessary to introduce a central network configuration and control. Underlayer networks are connected to the overlay network via a gateway. This gateway is used to transmit configurations commands to the underlayer network from the central network controller and to exchange data between overlay and underlayer networks. Particularly when data is exchanged between different operators, trustworthy communication between all parties in the network must be ensured \cite{RadioResource1}. With the introduction of 6G underlayer networks, application-oriented communication can be established, which can adapt to changing requirements in an organic and agile way by means of network control \cite{RadioResource2}.

\subsection{Network configuration}
\label{sec32}
Currently three distinct networks are planned. They share 100 MHz of spectrum between 3.7 and 3.8 GHz which can be licensed by the Bundesnetzagentur (BNetzA)\footnote{%
    BNetzA : \textit{Regionale und lokale Netze}, https://www.bundesnetzagentur.de/\\DE/Fachthemen/Telekommunikation/Frequenzen/OeffentlicheNetze/LokaleNe\\tze/lokalenetze-node.html (2020)%
} for local use. The spectrum will be split into two 20 MHz blocks and one 60 MHz block. The two smaller blocks are used for the underlayer networks. One for the \ac{urllc} of the controllers and the other one for sensor data communication. The two underlayer networks share the same master node which is bridging the connection to the overlay 6G network. The overlayer network is using the allocated 60 MHz for non mission critical communication with relaxed \ac{qos} parameters. 

The current network configuration is static, but for larger systems with more dynamic applications an automated network management is required. The master node of the underlayer network, which is always connected to the overlayer network, can request spectrum at a specific place from the network management entity. The current required spectrum at this place can be calculated and the decision to grant or reject the request can be made. Since the underlayer network is location specific with a small footprint more than one underlayer network could conceivably be operated in the same cell of the overlayer network.

\section{Use Case: Closed-loop communication for machine tools}
\label{sec4}
\subsection{Requirements for applications in manufacturing}
\label{sec41}

As described in Section \ref{sec23}, different requirements have to be met to enable the utilization of wireless communication technologies for manufacturing. Moreover, closed-loop machine tool control is the use case with the highest requirements regarding communication performance in terms of latency, jitter, and reliability regarding to 3GPP TR 22.804\cite{3GPP.072020}.

The requirements for the overall system can be summarized as follows:
\begin{itemize}
    \item Low-latency communication and time determinism: 0.5 to 10 ms end-to-end latency to enable closed-loop control
    \item Reliability and robustness: Low packet loss rate and network robustness to enable safety-critical applications (packet loss rate below $10^{-9}$)
    \item Integration to overlay network: Simultaneous deployment of different wireless communication technologies that are interconnected to enable the required scalability and flexibility of \ac{cpps}
\end{itemize}
To meet these above listed requirements the setup described in the following Section has been developed and evaluated. 

\subsection{Setup}
\label{sec42}
\begin{figure*}[ht]
    \centering
    \includegraphics[width=\textwidth]{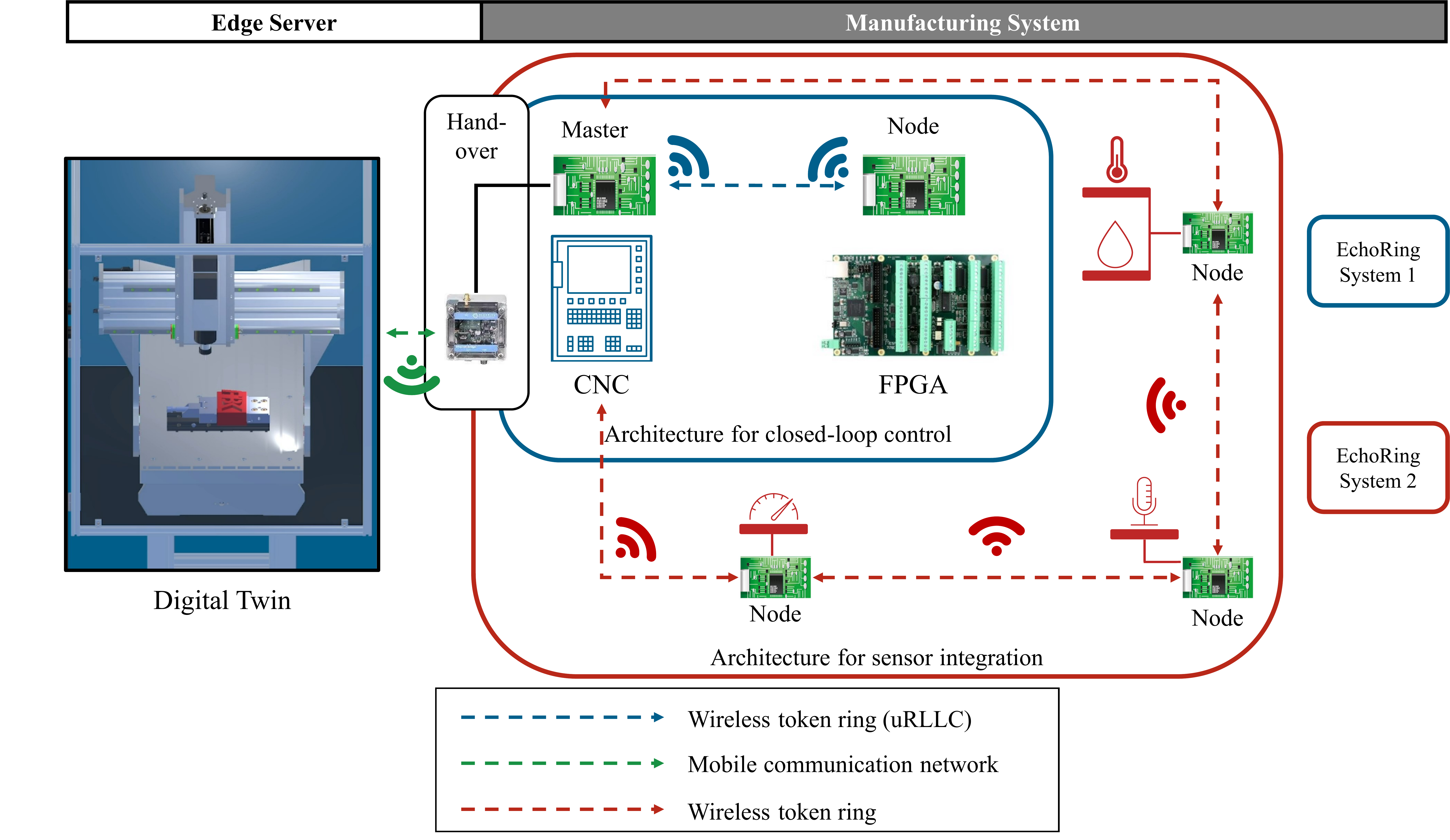}
    \caption{Implemented Setup}
    \label{fig41}
\end{figure*}

As shown in Figure \ref{fig41}, the implemented setup consists of different hardware and software components as well as different communication systems. Regarding the utilized hardware, a 3-axis milling machine tool and a CNC unit for machine tool control is deployed on the shop floor. The machine tool consists of actuators and sensors such as stepper motors, the tool spindle, limit switches and vibration sensors. Moreover, an Ethernet-capable FPGA (MESA 7i76E) to handle motion control is connected to the CNC. In addition, an edge server that is located near the shop floor is used to offload information of non-latency critical applications via a mobile communication network.  

The communication infrastructure for the underlayer network consists of two wireless token ring systems - so called EchoRing \cite{domb15} - for latency-critical and reliable communication and sensor integration (latency $<$ 2ms, packet loss rate down to $10^{-9}$)\footnote{%
   R3 Solutions: \textit{Data Sheet - Bridge E}, https://cta-redirect.hubspot.com/cta/redirect/4230617/a4178d6b-0a7d-4d4d-b3ac-a4548f41ca2b (2013-06)%
}. The wireless token ring systems enable communication between various components on the shop floor. Specifically, an \ac{urllc} ring between the CNC and FPGA is deployed to facilitate machine tool control, including PID control. Communication is based on UDP to enable low latencies. Another system with less rigorous requirements with support of up to eight devices is implemented for wireless sensor integration into the closed-loop system.

In addition, a 6G mobile communication network for non-time-critical communication is part of the communication system. The information between the two token ring systems are merged by a master device that connects and transfers them to the 6G mobile communication network.

On the software side, LinuxCNC is utilized as an open and adaptable CNC software. In addition, a digital twin of the machine tool is developed based on the Unity gaming engine, which enables monitoring, manual control, simulation and diagnosis of the process. The digital twin is running on the edge server and gathers information via the 6G network from sensors and the CNC unit. Thus, interfaces were developed to seamlessly integrate sensors into both the CNC system and the digital twin, enabling efficient data acquisition and utilization.

\subsection{Evaluation of functionality}
\label{sec43}

To evaluate the functionality of the proposed architecture based on the EchoRing system, a wired test setup is implemented to estimate the needed network performance characteristics regarding latency and jitter. The test setup is is utilizing \ac{nm} which is integrated into the Linux kernel. The kernel clock rate has been adapted to 1 kHz to enable deterministic delay emulation of 1 ms. The average packet size transmitted via UDP is around 80-159 bytes. 

The different configurations are shown in Table \ref{tab1}. The experiments started with the highest combination of the performance characteristics (latency = 5 ms, jitter = 0.3 ms). For each working configuration the operation of machine tool has been tested for one hour. If the operation was not interrupted due to following errors of the feedback control, the functionality can be considered as valid. In Table \ref{tab1}, the combinations that did not work are marked with an "x". For the combinations with "(\checkmark)", an adaption of the linux driver for initialization of the FPGA had to be done. The combinations that worked without any adaptions are marked with "\checkmark". 

The experiments show that the maximum latency to operate the machine tool is 3 ms with a the maximum jitter of 0.2 ms. However, driver adaptions, tuning of the PID controller, the watchdog, as well as the cycle times of the CNC of the machine tool is needed for stable operation of the system. Therefore, it can be concluded that the EchoRing solution meets the communication requirements as it provides time-deterministic, low-latency communication below 2ms. 

\begin{table}[]
\centering
\caption{Experimental results regarding network KPI for the CNC machine tool}
\label{tab1}
\begin{tabular}{ll|l|l|l|l|l|l}
                           &       \multicolumn{7}{c}{~~~~~~~~~~ Latency in ms}     \\
                           &      & 0.5 & 1   & 1.5 & 2   & 3   & 5 \\
                           \hline
\multicolumn{1}{c}{Jitter in ms} & 0.05 &  \checkmark   & \checkmark   & \checkmark   & \checkmark   & \checkmark   & x \\
\hline
                           & 0.1  & \checkmark   & \checkmark   & \checkmark   & \checkmark   & \checkmark  & x \\
                           \hline
                           & 0.15 & \checkmark   & \checkmark   & \checkmark   & \checkmark   & \checkmark   & x \\
                           \hline
                           & 0.2  & \checkmark   & (\checkmark) & (\checkmark) & (\checkmark) & (\checkmark) & x \\
                           \hline
                           & 0.3  & x   & x   & x   & x   & x   & x
\end{tabular}
\end{table}

\subsection{Benefits and challenges}
\label{sec44}
The implemented architecture leads to different benefits. First, it enables flexible and scalable manufacturing systems due to meeting various requirements simultaneously regarding the communication performance. Even wireless communication between machine tools and CNC units for machine tool control can be realized. Moreover, serveral EchoRing systems can be deployed and integrated simultaneously for different network requirements. Due to  integration of the underlayer networks into a mobile communication network, scalability - especially for data intensive, non-latency critical use cases - is enabled. Second, the wireless communication architecture facilitates retrofitting of numerous equipment (e.g. sensors, actors, computing units) in existing manufacturing systems or for machine tools. Third, the system enables real-time communication and time determinism in a wireless communication architecture for manufacturing. This  leads to new use cases in manufacturing systems e.g. for real-time control, virtualized SPS or real-time diagnosis with retrofit sensors .

By introducing an efficient network management in Section \ref{sec3}, which takes over the coordination of the underlayer networks among each other as well as with the overlayer network, first approaches for a successful implementation of these new use cases were introduced. However, there are several challenges that need to be addressed. One challenge is the timing synchronization when integrating several EchoRing systems with different cycle times into the closed-loop of a CNC system. Another challenge is the performance of the wireless machine tool control system in a real-world setting. Therefore, testing on the shop floor is necessary. Factors such as harsh environmental conditions, including massive machine housing and covered line of sight, need to be investigated. Moreover, the deployment of multiple different wireless systems operating in the same manufacturing systems could lead to interferences and thus to inhibited scalability. 

\section{Conclusion and future work}
\label{concl}
In this paper, we have presented a concept for 6G underlayer networks in the scope of robust and low-latency communication. While wireless transmission offers greater flexibility and ease of integration, the introduced concept of underlayer networks ensures the reliability, low latency and security required for seamless and efficient communication in highly interconnected and autonomous manufacturing systems. We introduced a underlayer network concept for 6G in manufacturing, which offers a robust and efficient communication infrastructure, enabling wireless connectivity, flexibility, and adaptability in a manufacturing environment. By integrating the idea of private networks, network slicing and spectrum sharing, a new management system is required to coordinate and route traffic across multiple networks. The underlayer network concept is put in the context of a manufacturing scenario. This architecture enables wireless closed-loop control of machine tools and seamless data transmission across different networks. It leverages the advantages of cellular communication networks, offering robust communication through a higher degree of determinism. The underlayer networks can be configured and controlled centrally, ensuring trustworthy, low-latency communication between all parties in the network. The concept of underlayer networks provides an agile and application-oriented communication solution that can adapt to changing requirements and support future 6G communication systems.

For future work, a more detailed examination of the introduced architecture in a dynamic manufacturing environment is planned, where the specific tasks of network control in non-static scenarios will be investigated and evaluated. In addition, the architecture is currently in a implementation phase. The setup will be subject of ongoing and future research, e.g. for determining the manufactured part quality with wireless closed-loop control.

\section{Acknowledgment}
The authors acknowledge the financial support by the German \textit{Federal Ministry for Education and Research (BMBF)} within the project Open6GHub \{16KISK004\}.

{%
\printbibliography%
}%

\end{document}